\documentclass{article}
\pdfoutput=1
\usepackage{graphicx}
\usepackage{amssymb}

\begin{document}

\begin{center}

{\Large {\bf Finite Energy Sum Rules with Legendre Polynomial Kernels\footnote{Work supported by Spanish MICINN under grant FPA2011-23596, FPA2014-54459-P, SEV-2014-0398 and GVPROMETEO 2010-056.

Talk given at 18th International Conference in Quantum Chromodynamics (QCD 15,  30th anniversary),  29 june - 3 july 2015, Montpellier - FR }
 }}

\vspace{.5cm}

\baselineskip 14pt
{\large J. Bordes\footnote{bordes@uv.es}, J. A. Pe\~{n}arrocha\footnote{jose.a.penarrocha@uv.es.
Speaker, corresponding author.
 }}
\\
{\it Theoretical Physics Department-IFIC (Universitat de Valencia and CSIC). \\
 C. Dr. Moliner 50, E-46100 Valencia Spain}
\\
and \\
{\large  Michael J. Baker\footnote{micbaker@uni-mainz.de}}
\\
{\it  PRISMA Cluster of Excellence \& Mainz Institute for Theoretical Physics, Johannes Gutenberg University, 55099 Mainz, Germany}

\end{center}

\begin{abstract}
In this note we report about a method to deal with finite energy sum rules.
With a reasonable knowledge of the main resonances of the spectrum, the
method guarantees that we can find a nice duality matching between the low
energy hadronic data and asymptotic QCD at high energies.
\end{abstract}


\vspace{.5cm}
\hrule
\vspace{.5cm}

\section{Finite Energy Sum Rule}

As a general definition, we can say that a Finite Energy Sum Rule (FESR) is
an equation that identifies a QCD theoretical calculation along a finite
region of energy in the physical strong interacting spectrum with the
experimental data in this corresponding energy region.

\subsection{Two point correlator.}

Our theoretical object is the vacuum expectation value of the time ordered product of two
currents ($j \left( x\right) $) corresponding to a particular channel (labeled by $\Gamma$) of strong interacting particles:

\begin{equation}
\Pi ^{\Gamma }\left( q\right) =i\int d^{4} x
\, \, e^{iqx} \,
\langle \Omega
\left\vert T\left\{ j^{\Gamma }\left( x\right) j^{\Gamma }\left( 0\right)
\right\} \right\vert \Omega \rangle   
\label{correlator}
\end{equation}

This correlator is related to an observable quantity ($\rho^{\Gamma}$), either scattering or decay,  through the imaginary part of its analytic structure. The Optical Theorem provides the following relationship:

\begin{equation}
\frac{1}{\pi}\, \mathrm{Im}\, \Pi^{\Gamma}\backsim\rho^{\Gamma}(\Omega
\rightarrow \mathrm{hadrons})\label{opticalTh}
\end{equation}

Applying this equation to QCD the first problem is that we cannot perform calculations in all ranges of energy. Namely, asymptotic QCD in the low
energy region where hadron resonances appear, is far from being accessible with the present techniques (aside of lattice calculations). Therefore, in order to relate
low energy properties of resonances (such as masses or decay constants) with the high energy QCD parameters (quark masses, strong coupling...) we need to
introduce appropriate techniques in the correlator to make possible the
matching between the two regime of energies. A master equation to achieve this target
is given by the Cauchy's Theorem of analytic functions.

\subsection{Cauchy's Theorem}

\begin{figure}[htb]
\vspace{9pt}
\begin{center}
\fbox{
\includegraphics[scale=0.9]{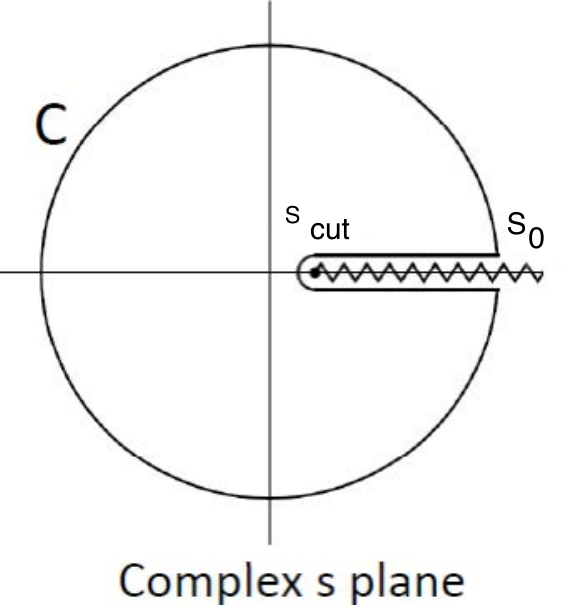}
}
\end{center}
\caption{Cauchy's Theorem on the $s$ plane}
\label{cauchy}
\end{figure}

For an analytic function $\Pi^{\Gamma}\left(s\right)$ in the $s$ complex
plane with a real cut starting in $s_{\mathrm{cut}}$, Cauchy's Theorem states:
\begin{equation}
\frac{1}{2\pi i} \,
{\displaystyle\oint\limits_{\mathrm{C}}}
\, \Pi^{\Gamma}\left(s\right)  ds \, = \, 0
\label{cauchyTh}
\end{equation}
C is a closed path surrounding the real cut (figure \ref{cauchy}).

Then, if we introduce an arbitrary entire function $K\left(s\right)  $ we can also write

\begin{equation}
\frac{1}{2\pi i}
{\displaystyle\oint\limits_{\mathrm{C}}}
K\left(  s\right) \,  \Pi^{\Gamma}\left(  s\right)  ds \, = \, 0
\label{cauchyKs}
\end{equation}

Being $\Pi^{\Gamma}\left(s\right)$ real in the real axes outside the cut we can use Schwarz Reflection Principle and write:

\[
\! \! \! \! \! 
 \frac{1}{\pi}\int_{s_{\mathrm{cut}}}^{s_{0}} \! \! \! \!  \! \!   K\left(s \right) \,  \mathrm{Im} \, \Pi^{\Gamma}\left(s\right)  ds
=\frac{-1}{2\pi i}\, \int_{\! \!  \left\vert s\right\vert \,  = \, s_{0}} \! \!  \! \!  \! \! K\left(s\right) 
\Pi^{\Gamma}\left(s\right)   ds
%\label{CauchyCut}
\]

In this relation we have split the contribution of the integral along the
closed path $C$ into two parts (figure \ref{cauchy}). In the l.h.s. the integration runs along the cut, 
including the low energy region starting at $s_{\mathrm{cut}}$, and in the r.h.s. the integral is performed 
along the circle of radius $s_{0}$
chosen in the high energy region. Along the cut we can use the 
experimental data information and in the circle we can calculate the
correlator using asymptotic QCD. In this way we match theory with experiment.

The confrontation of theory and experiment by means of Cauchy's Theorem yields
what we call the finite energy sum rule
\begin{equation}
 \frac{1}{\pi}\int_{s_{\mathrm{phys}}}^{s_{0}}
 K\left(  s\right) \, \mathrm{Im} \, \Pi_{\mathrm{Data}}^{\Gamma}\left(  s\right)  ds  
 \approx\,  \!\! -\frac{1}{2\pi i}\int_{\left\vert s\right\vert =s_{0}}
K\left(s\right) \,  \Pi_{\mathrm{QCD}}^{\Gamma}\left(  s\right)  \, ds
\label{cauchyDat}
\end{equation}
where $s_\mathrm{phys}$ stands for the physical threshold.

Taking into account that the function $\Pi_{QCD}^{\Gamma}\left(  s\right)$
has its own analytic structure, with a QCD cut in the real axes starting at
$s_{QCD}$, we can substitute the r.h.s. of the former equation by an integral of the imaginary part
along the cut, thus an equivalent relation is:
\begin{equation}
\frac{1}{\pi}\int_{s_{\mathrm{phys}}}^{s_{0}}K\left(s\right) \,  \mathrm{Im}\,  
\Pi_{\mathrm{Data}}^{\Gamma}\left(  s\right)  \,  ds 
 \approx \,  \frac{1}{\pi}\int_{s_{QCD}}^{s_{0}}K\left(  s\right)
\mathrm{Im}\Pi_{\mathrm{QCD}}^{\Gamma}\left(  s\right)  \,  ds 
\label{CauchyQCDth}
\end{equation}

Here we have to face the following problem. From the experimental data we
usually know the physical spectrum near the physical threshold,  where
resonances appear, whereas at high energies we do not have 
information of the final multiparticle product spectrum. In order to match this information with the 
results of QCD at high energies we introduce a new parameter in the sum rule, that we call $s_{\mathrm{eff}}$, 
above which we are allowed for the substitution of experimental data by QCD in the physical cut.

\subsection{The value of $s$ effective ($s_{\mathrm{eff}}$)  }

Let us define the effective threshold satisfying
\begin{equation}
\mathrm{Im}\, \Pi_{\mathrm{Data}}^{\Gamma}\left(  s\right)    \approx
\mathrm{Im}\Pi_{\mathrm{QCD}}^{\Gamma}\left(  s\right) \quad
\mathrm {for } \, s   \geqslant s_{\mathrm{eff}}
 \label{duality}
\end{equation}
and
\begin{equation} \frac{1}{\pi}\int_{s_{\mathrm{phys}}}^{s_{\mathrm{eff}}}K\left(  s\right) \mathrm{Im}  \,
\Pi_{\mathrm{Data}}^{\Gamma}\left(  s\right)  ds 
 \approx\frac{1}{\pi}\int_{s_{QCD}}^{s_{\mathrm{eff}}}K\left(s\right)  \, 
\mathrm{Im}  \, \Pi_{\mathrm{QCD}}^{\Gamma}\left(  s\right)  ds
\label{duality Integral}
\end{equation}
This last relation does not follow from (\ref{CauchyQCDth}) just from the approximation of integrands (\ref{duality}). We need
also to choose suitable ``kernels" $K\left(  s\right)$.
As a result we face with two problems: how to choose the appropriate
kernel and how to fix the value for $s_{\mathrm{eff}}$ to get a good approximation in equation (\ref{duality Integral}).

\begin{figure}[htb]
\vspace{9pt}
\begin{center}
\fbox{
\includegraphics[scale=0.55]{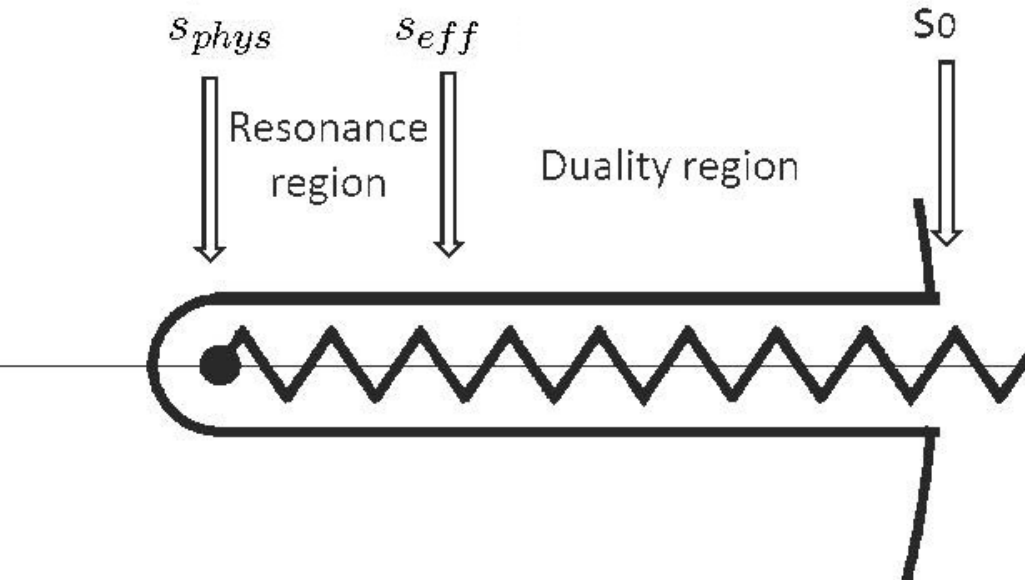}
}
\end{center}
\caption{Physical cut}
\end{figure}

\section{Choosing the appropriate Kernel}

For sake of comparison we will focus our discussion in two kind of kernels: the
exponential kernel and the Legendre Polynomial kernels, which are the object of this note. 

\subsection{Exponential Kernel: SVZ-Laplace Sum Rule}

The exponential kernel, that was first introduced in \cite{SVZ}, enhances the resonance region in the energy interval $\left[s_{\mathrm{phys}},s_{\mathrm{eff}}\right]$  and is given by
$$
K\left(  s,M^{2}\right)  =e^{-s/M^{2}}\label{exponential}
$$
This falling down of the exponential kernel works in such a way that, on both the experimental data and the asymptotic QCD integrals,
we can neglect the part of the integral in the interval $\left[s_{\mathrm{eff}}, s_{0}\right]$
even in the limit $s_{0}\rightarrow\infty.$ 

Then, with this kernel the Sum Rule (\ref{duality Integral}) becomes
\begin{equation}
\frac{1}{\pi}\int_{s_{\mathrm{phys}}}^{s_{\mathrm{eff}}}e^{-s/M^{2}} \, \mathrm{Im} \,
\Pi_{\mathrm{Data}}^{\Gamma}\left(  s\right)  ds
 \approx\frac{1}{\pi}\int_{s_{QCD}}^{s_{\mathrm{eff}}}e^{-s/M^{2}} \,\mathrm{Im} \,
\Pi_{\mathrm{QCD}}^{\Gamma}\left(  s,m\right)  ds
\label{exponentialSR}
\end{equation}

In the r.h.s. of (\ref{exponentialSR}) we have  introduced a QCD mass parameter ($m$) to be
estimated. The prediction of $m$ is done at some particular $M$ provided that there exists
a neighbourhood of $M$ where this prediction is constant.
Notice that  $M$ fixes the slope
of the exponential kernel in the resonance region, i.e. the amount of
enhancement or suppression we introduce with the kernel. 

\subsection{Legendre Polynomial (LP) Kernels}

This set of kernels, that was first successfully used in the calculation of the heavy quark masses \cite{Charm2001,Bottom2003}, 
is introduced defining a set of n-degree orthogonal polynomials $P_{n}(s)$ in the energy interval
$s\,\in\,[s_{\mathrm{eff}},s_{0}]$
with a normalization condition at the
lowest lying resonance, $s=M_{R}^{2}$, as follows:
\begin{eqnarray}
&& \int_{s_{\mathrm{eff}}}^{s_{0}}s^{q}\,P_{n}(s)\,ds  
=0,~\ q=0,,1...,n-1  \nonumber  \\
&& P_{n}(M_{R}^{2})   =1
\label{polynomialCondi}
\end{eqnarray}

These polynomials are related to the ordinary Legendre polynomials
$\mathcal{P}_{n}(x)$ in the interval $x\,\in\,[-1,1]$:
\[
\!\!\!\!
P_{n}(s)\,=\,\frac{\mathcal{P}_{n}\left(  x(s)\right)  }{\mathcal{P}
_{n}\left(  x(M_{R}^{2})\right)},  \quad
x(s)\,=\,\frac{2s\,-\,(s_{0}+s_{\mathrm{eff}})}{s_{0}-s_{\mathrm{eff}}}
\]

We quote here some of the Legendre Polynomials we use:
\begin{eqnarray}
 \mathcal{P}_{2}\left(  x\right) & = & \frac{1}{2}(3x^{2}
-1), \nonumber \\
\mathcal{P}_{3}\left(  x\right)  & = & \frac{1}{2}(5x^{3}-3x), \nonumber \\
\mathcal{P}_{4}\left(  x\right)  & = & \frac{1}{8}(35x^{4}-30x^{2}
+3), \nonumber \\
\mathcal{P}_{5}\left(  x\right)  & = & \frac{1}{8}(63x^{5}-70x^{3}
+15x).
\label{legendrePolynomial}
\end{eqnarray}

Notice that the oscillatory nature of the polynomials cancels the unknown data in the interval 
$ \left[  s_{\mathrm{eff}},s_{0}\right]$  more efficiently than in the exponential case.

We quote here some of the Legendre Polynomials we use:
\begin{eqnarray}
& \frac{1}{\pi}\int_{s_{\mathrm{phys}}}^{s_{\mathrm{eff}}}P_{n}\left(s;s_{\mathrm{eff}},s_{0}\right)
\, \mathrm{Im} \, \Pi_{\mathrm{Data}}^{\Gamma}\left(s \right)  ds 
\nonumber \\
& \approx\frac{1}{\pi}\int_{s_{QCD}}^{s_{\mathrm{eff}}}P_{n}\left(s;s_{\mathrm{eff}},s_{0}\right)  \,  \mathrm{Im} \, \Pi_{\mathrm{QCD}}^{\Gamma}\left(s,m\right)
ds
\label{legendreSR}
\end{eqnarray}
where  $P_{n}\left(s;s_{\mathrm{eff}},s_{0}\right)$ are defined in the range
$s \, \in \, \left[  s_{\mathrm{eff}},s_{0}\right]$  according to (\ref{polynomialCondi}).

Now the method to determine the QCD parameter $m$ within this sum rule works as follows:

\begin{itemize}
\item As a first step we study in equation (\ref{legendreSR}) the stability of our prediction of $m$ with $s_{0},$ taking
$s_{\mathrm{eff}}$ around the continuum physical threshold, beyond the resonance region.
\item Then we adjust the value for $s_{\mathrm{eff}}$ by demanding optimal stability, namely that the function
$m\left(s_{0}\right)$ remains constant in a plateau as large
as possible. 
\item The estimate for $m$ is taken for the optimal stability values of the parameters $s_0$ and $s_{\mathrm{eff}}$.
\item Finally, we check the convergence of the result with the degree of the polynomial ($n$).
\end{itemize}

To see how this sum rule method works in a particular calculation, we describe
in the next section the determination of the $B$ decay constant ($f_B$).

\section{Legendre Polynomial Kernel in the $f_{B}$ determination.}

\begin{figure}[htb]
\vspace{9pt}
\begin{center}
\includegraphics[scale=0.3]{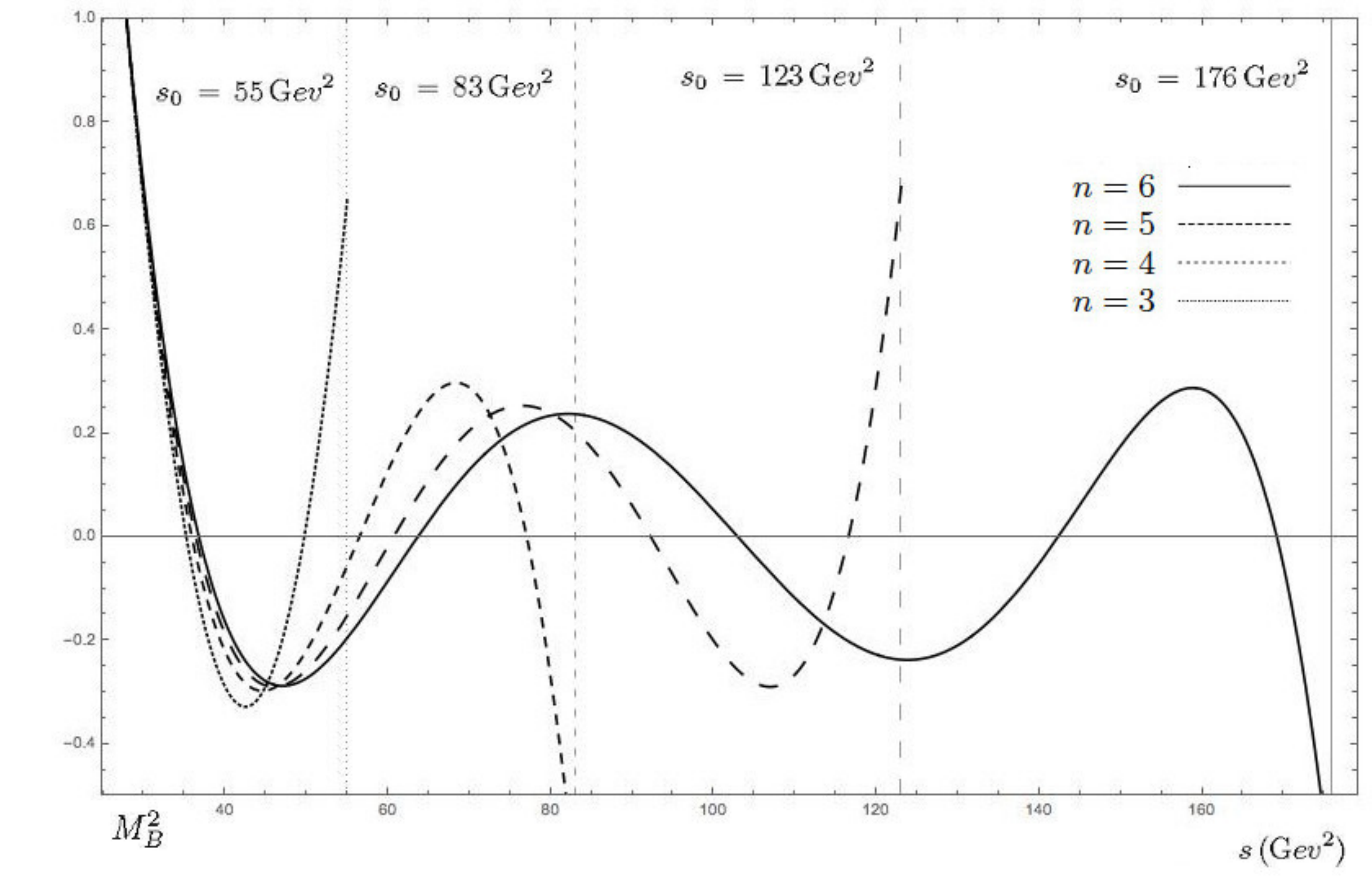}
\end{center}
\caption{Legendre polynomials for $f_{B}$ with threshold $s_{phys}=29.9 \, {\mathrm Gev}^2$  (see equation (\ref{legendrePolynomial}))}
\label{LPstability}
\end{figure}

Taking the pseudoscalar correlator for the pair of quarks $ub$ and assuming the dominance of the lowest lying resonance ($ B $ pole) we take:
\begin{equation}
\frac{1}{\pi}\, \mathrm{Im} \, \Pi_{\mathrm{Data}}^{\Gamma}\left(  s\right)  \, = \, f_{B}
^{2} \, M_{B}^{4}\, \delta\left(  s-M_{B}^{2}\right), \label{fBpole}
\end{equation}
being the decay constant $f_B$ the unknown in equation (\ref{legendreSR}):
\begin{equation}
\! \! \! \! \! \! \! \! \! \! \! 
f_{B}^{2}\left( s_{0}\right)  \approx
\frac{1}{\pi M_{B}^{4}}
\int_{s_{QCD}}^{s_{\mathrm{eff}}}  
\! \! \! \! \! \! 
P_{n}\left(s;s_{\mathrm{eff}},s_{0}\right) 
\mathrm{Im}\,\Pi_{\mathrm{QCD}}^{\Gamma}\left(  s\right)  ds
\label{fBsr}
\end{equation}
The dependence of $f_{B}$ in $s_{0}$ is implicit in the definition of the
polynomial $P_{n}\left(s;s_{\mathrm{eff}},s_{0}\right) $ (\ref{polynomialCondi}). Choosing $s_{\mathrm{eff}}$ at the
continuum physical threshold, $s_{\mathrm{eff}}=\left(  M_{B}+2m_{\pi}\right)  ^{2}=29.9 \, GeV^{2}$ as in \cite{B2004,D2005},
the polynomials that give stable results for different degrees $n$ are depicted in figure \ref{LPstability}.

We see that the higher n we choose, the larger values for $s_{0}$ we need to find stability,
however the slope of the polynomial remains roughly constant in the threshold
region. Again we find, as in the exponential case, that the amount of enhancement of  the
resonance region is a fundamental issue in the sum rule method.

\begin{figure}[htb]
\vspace{9pt}
\begin{center}
\includegraphics[scale=0.42]{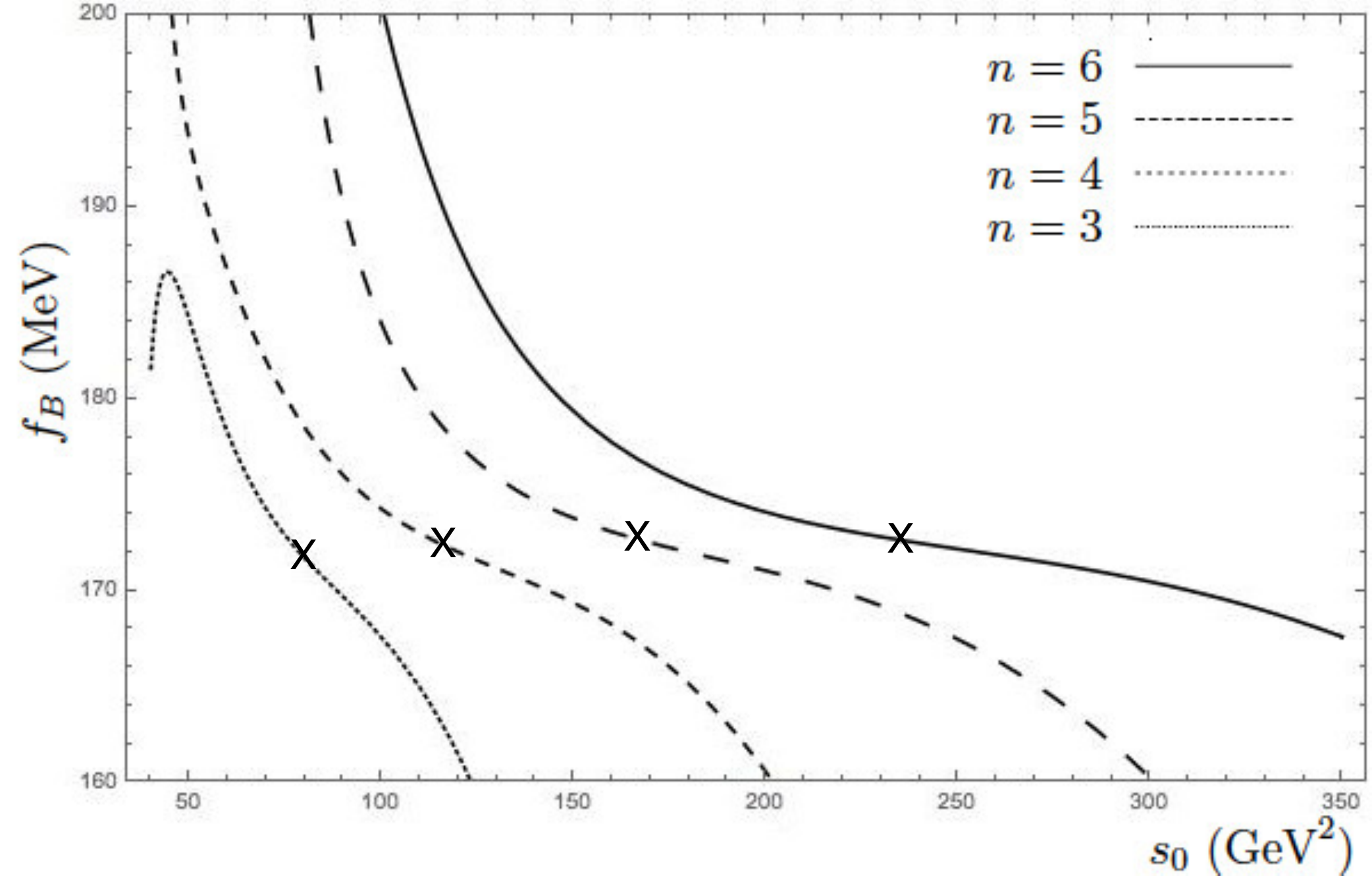}
\end{center}
\caption{$f_{B}$ results for different polynomials at the physical threshold $s_{phys}=29.9 \, {\mathrm Gev}^2$ (see text).}
\label{fBsphys}
\end{figure}

Also looking at the stability regions for different n degree of the LP kernel we appreciate in figure \ref{fBsphys}
that increasing the degree of the polynomial the plateau gets wider and the results, taken at the
inflexion points, show a very good convergence with the polynomial degree, the result going to the value $f_B \sim 175 \, \mathrm{Mev}$.

\begin{figure}[htb]
\vspace{9pt}
\begin{center}
\includegraphics[scale=0.38]{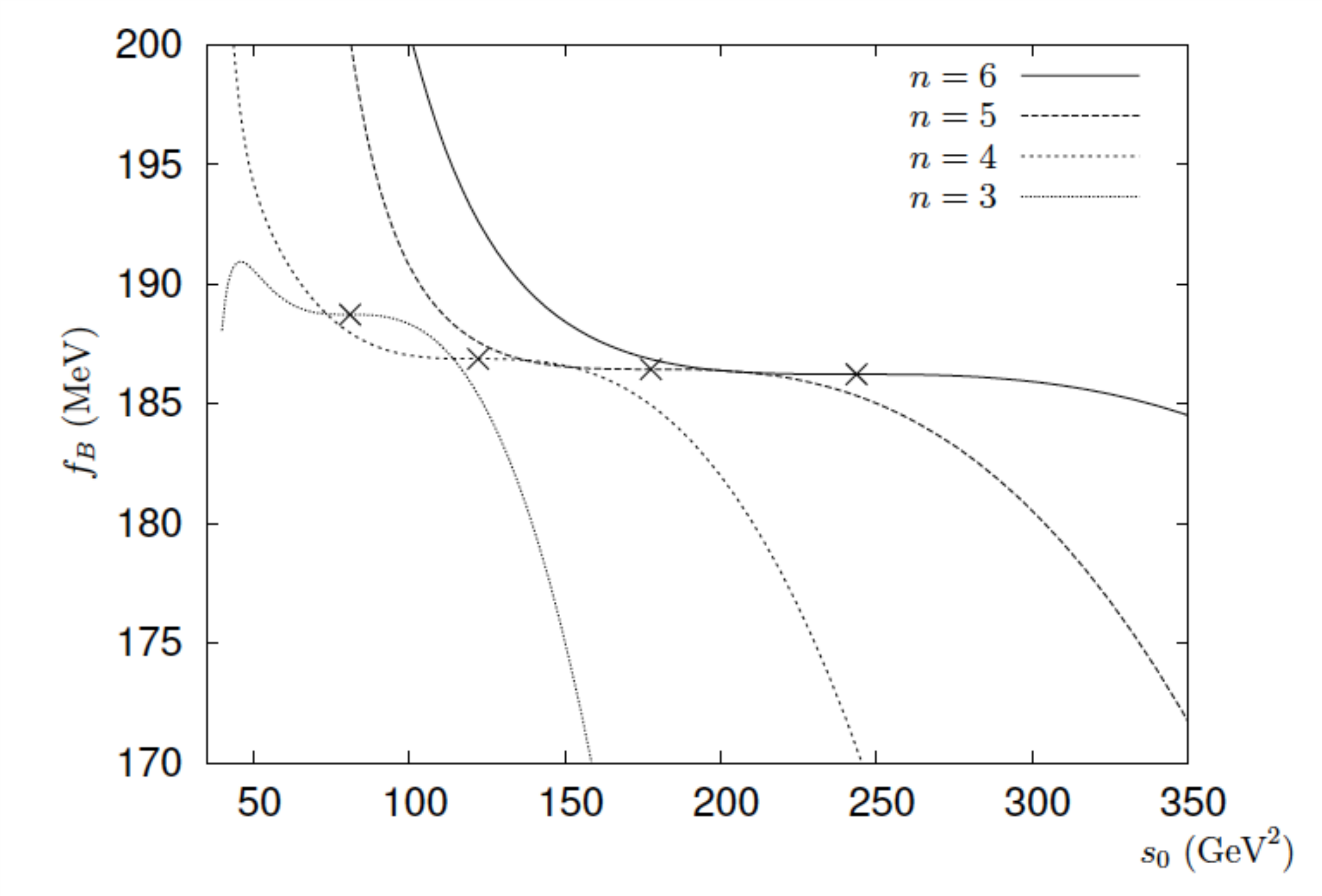}
\end{center}
\caption{$f_{B}$ results for different polynomials at optimal $s_{eff}=31.3 \, {\mathrm Gev}^2$ (see text).}
\label{fBseff}
\end{figure}

Next, instead of fixing $s_{\mathrm{eff}}$ at the continuum physical threshold, we
determine it by demanding optimal stability \cite{B2014}. In our case we will achieve this by imposing that the first derivative should also vanish at the inflexion point. Choosing this way to determine $s_{\mathrm{eff}}$ we improve substantially the stability region (as it is shown in figure \ref{fBseff}) with a result for the decay constant slightly higher ($f_B \sim 186 \, \mathrm{Mev}$) which is constant in a wider range of $s_0$. The value of optimal stability that we find for $s_{\mathrm{eff}}$ is not far from the physical continuum threshold, as one could anticipate, although the difference is crucial for stabilizing the result. This feature is one of the main motivations to
trust the sum rule method with LP kernels.

A further advantage of this method is that the contribution of the sum rule
integral in the region $\left[s_{\mathrm{eff}},s_{0}\right]$ is really tiny. As a matter of fact we have compared the QCD integration pieces  in the $f_{B}$ calculation obtaining:
\[
\frac{\int_{s_{\mathrm{eff}}}^{s_{0}}P_{n}\left(s;s_{\mathrm{eff}},s_{0}\right) \,
\mathrm{Im}\Pi_{\mathrm{QCD}}^{\Gamma}\left(  s,m\right)  ds}{\int_{s_{QCD}
}^{s_{\mathrm{eff}}}P_{n}\left(  s;s_{\mathrm{eff}},s_{0}\right)  \, \mathrm{Im}\Pi
_{\mathrm{QCD}}^{\Gamma}\left(  s,m\right)  ds} \,\thicksim \, 0.005
\]
The same would happen in the experimental data integration provided we take $s_{\mathrm{eff}}$
beyond the resonance region.
Obviously the lack of precise information introduces a systematic uncertainty of
the method which is beyond our control. Nevertheless, we expect this uncertainty to be small, since even the
resonances in the continuum region are substantially suppressed by the
polynomials with respect to the lowest lying resonance (see figure \ref{LPstability}).

\section{Legendre Polynomial Kernel in the  determination of the strange quark mass}

The calculation of the light quark masses from the pseudoscalar current is a bit
cumbersome due to the poor convergence of the QCD correlator with
respect to the strong coupling. We give here some preliminary results using the LP
sum rule method.

\begin{figure}[htb]
\vspace{9pt}
\begin{center}
\includegraphics[scale=0.8]{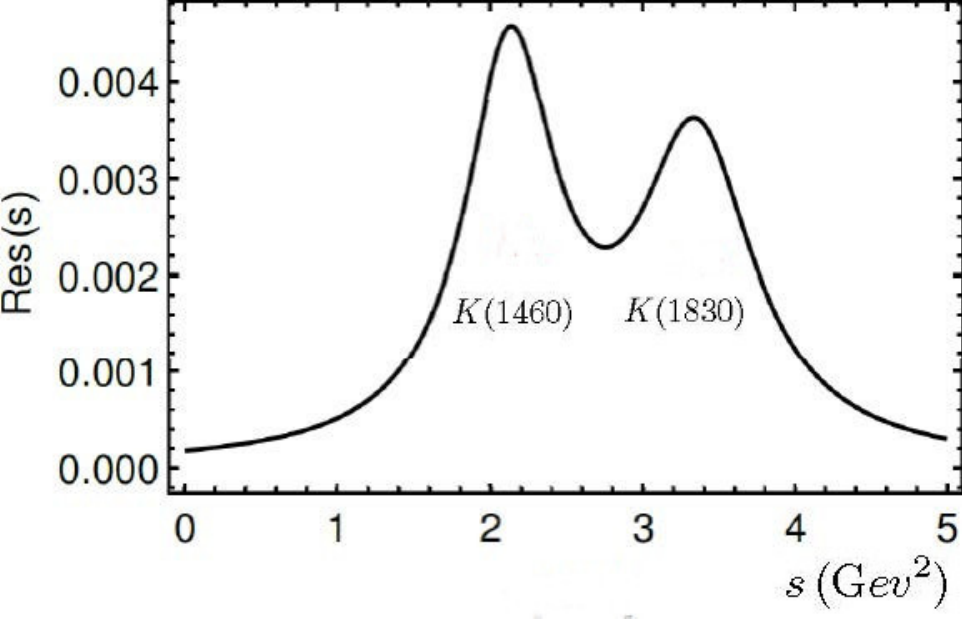}
\end{center}
\caption{Resonance model in the Kaon channel (the kaon pole is omitted).}
\label{Kresonances}
\end{figure}

For the strange quark mass we  use in the spectral function, aside from the Kaon pole, the contribution
of the $K(1460)$ and $K(1830)$ resonances that significantly improves the convergence of the results. 
In order to perform the integration of the experimental side we use a Breit-Wigner model, as depicted in figure \ref{Kresonances}.

\begin{figure}[htb]
\vspace{9pt}
\begin{center}
\includegraphics[scale=0.45]{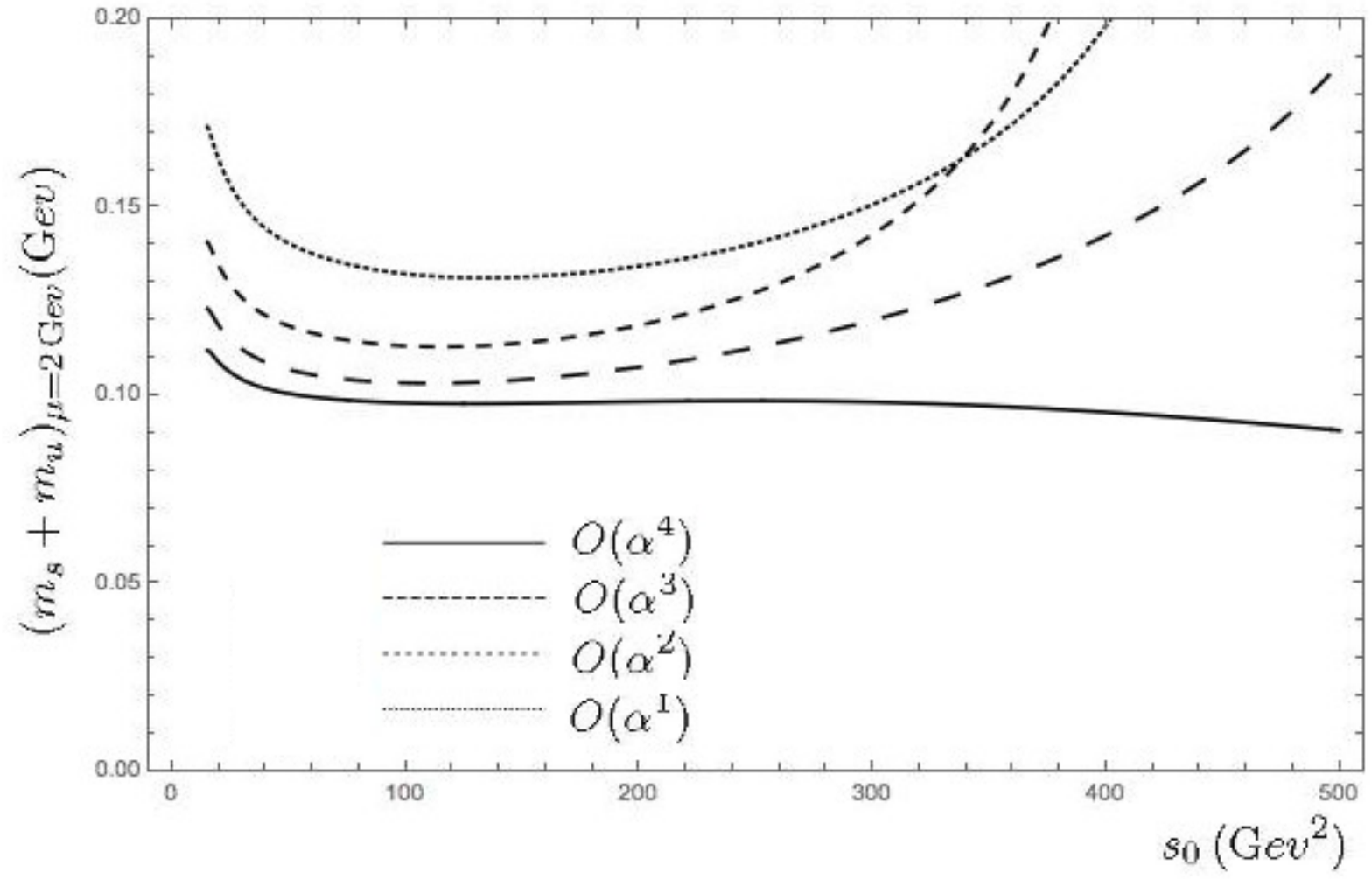}
\end{center}
\caption{$m_{s}+m_u$  results at different orders of the strong coupling constant
with a 5th degree polynomial ($s_{eff}=3.6 \, {\mathrm Gev}^2$).}
\label{ms}
\end{figure}

To have a flavour of the results obtained we present in figure \ref{ms} the stable values for the quark masses obtained with  the 6th order Legendre polynomial. The stability regions of the mass with the value of $s_0$ is apparent,
presenting again a nice plateau around the stability points. Results for different orders in the strong coupling constant show a fairly good convergence of the results for the masses. This convergence is worse than in the light-heavy quarks system calculations and further investigation is under way.

Aside from this consideration, what is important to stress here concerning the LP sum rule method, is that
the value for $s_{\mathrm{eff}}$ is determined by optimal stability and it is located, as expected, just
after the resonance region where one expects that the experimental results approach to some smooth function of the energy. On the other hand we have checked that the result has a very good convergence
when the degree of the polynomial is increased.

\section{Conclusions}

To summarize, we have reviewed the main features of the FESR method with LP kernels. The main
advantages that we find are:

\begin{itemize}
\item The LP kernels eliminate very efficiently the contribution of
asymptotic QCD and the experimental data in the interval $\left[ s_{\mathrm{eff}},s_{0}\right] $ of the
Sum Rule integral.
\item The LP kernels are easy to integrate with asymptotic QCD in the circle $\left\vert
s\right\vert =s_{0}$ and one does not need to extract the imaginary part from the QCD correlator.
\item The method is able to determine $s_{\mathrm{eff}}$  in a systematic way, which provides a better stability of the results (figure \ref{fBseff}).
\item The final results show nice plateaus with $s_{0}$ as well as a good convergence with the degree of the polynomial. 
\item The slope of the LP near the threshold approaches to a constant when increasing the degree
of the polynomial.
\item First used in the calculation of the heavy quark masses \cite{Charm2001,Bottom2003} we have used it recently in the calculation of the meson decay constants in the heavy-light quark systems \cite{B2004,D2005,B2014} and work in the pion and kaon channels is in progress \cite{Strange quark mass (preliminary)}.
\end{itemize}

\end{document}